\documentclass[]{article}
\usepackage{hyperref}
\title{Quantum Gravity for Dummies}
\author{Deepak Vaid \\ NITK, Surathkal \\ deepak@nitk.edu.in \\ \date{\today} }

\begin{document}

\maketitle

\begin{abstract}
I have been asked to write brief, gentle introduction to the basic idea behind the field of ``quantum gravity'' in 1500 words or less. Doing so appears to be almost as great a challenge as coming up with a consistent theory of quantum gravity. However, I will try. Disclaimer: \emph{The views expressed in this article are my own and do not represent the consensus of the quantum gravity community}.
\end{abstract}


\section{Semantics} 
\label{sec:semantics}

To get some idea of quantum gravity it is helpful to look at the meaning of those two words separately. The word ``quantum'' refers to the theory of quantum mechanics which was developed in the 20th century and which incorporates various seemingly paradoxical properties of light and matter such as wave-particle duality, the uncertainty principle, and the probabilistic nature of measurements into a coherent theoretical framework. Quantum Mechanics undergirds the basic technological framework of the world we live in. Without quantum mechanics we would not understand how semiconductors work and without semiconductors we would not be able to build the integrated circuits that power most modern electronic devices. Without QM we would not have NMR machines, CT scans, electron microscopes, superconductors or superfluids. Many of the theoretical implications of quantum mechanics have been realized outside the laboratory and are embedded in the many layers of technology that surround us. Many more of its implications - quantum computation and teleportation for e.g. - are still in the laboratory, not yet mature enough to venture out into the real world.

The word ``gravity'' refers not to Newton's theory of gravitation, but to Einstein's theory of general relativity which superseded the Newtonian conception of gravity. In the Newtonian framework, gravity was thought of as a force which acted between any two massive bodies in a manner given by the well-known formula: $F = G m_1 m_2/r^2$. By 1905 the Newtonian concept of motion, grounded in the postulates of absolute time and absolute space, had dissolved to give way to the relativistic approach developed by Einstein. This new framework was called the theory of ``special relativity''. The word ``special'' here stands for the fact that this theory is restricted to describing the behavior of bodies in inertial frames of reference - i.e. those that are non-accelerating. Lifting this restriction, involves a generalization of the theoretical framework to allows a description of motion in both inertial (non-accelerating) and non-inertial (accelerating) frames of reference. This is the origin of the prefix ``general'' in the term ``general relativity''. GR is, simply speaking, the extension of the special theory from inertial observers to arbitrary observers (or frames of reference).

Einstein's great insight was the realization that physics in a uniformly accelerating frame of reference is indistinguishable from physics in a constant gravitational field such as that experienced by observers close to the surface of the earth. This allowed him to construct a theory which described gravitation, not as a force acting between massive bodies, but as the manifestation of the geometry of the spacetime surrounding any given configuration of matter. Special Relativity had abolished the notion of an ``absolute space'' or ``absolute time'' in which all bodies executed their motions, in exchange for a framework where only the relative motion between two bodies was relevant. General Relativity went even further and proclaimed that the fabric of geometry in which all matter is embedded can is distorted by the presence of matter and that this distortion is what is perceived by us to be the ``force'' of gravity. The implications of this new understanding are vast and were only gradually discovered. Two of the new phenomena that are possible in General Relativity (and of whose existence reliable observational evidence exists) are black holes - regions of spacetime where the concentration of matter is great enough that even light cannot escape (Stephen Hawking's recent sensational pronouncement \cite{Hawking2014Information} notwithstanding) and the big bang - the primordial origin of the Universe itself.

\section{Unifying Perspective} 
\label{sec:unification}


The primary guiding philosophy in the development of physics over the past two centuries has been the idea of \emph{unification}. James Clerk Maxwell unified the theories of Faraday, Ampere and others about the electric and magnetic properties of matter into a single framework known as ``electromagnetism''. Maxwell's theory, among other achievements, predicted the existence of oscillations of electromagnetic fields whose predicted speed matched that of the speed of light, leading to the identification of light with waves of the electromagnetic field. Einstein's theory of Special Relativity served to reconcile Maxwell's theory of the electromagnetic field with the motions of material bodies.

In its infancy Quantum Mechanics was a brand new science, born out of the failures of classical physics. The initial, crude identification of electromagnetic fields with quantum mechanical entities known as \emph{photons}, led in the decades between WWI and WWII to the development of Quantum Electro Dynamics (QED) which provides a unified description of electromagnetic fields in a fully quantum setting. Following WWII rapid discoveries of a whole zoo of new elementary particles led theorists to postulate the existence of two more forces of Nature - in addition to the already known electromagnetic and gravitational forces - known as the ``weak'' and ``strong'' forces respectively. Theoretical work by giants of 20th century physics, such as Feynman, Gell-Mann and t'Hooft among many others, led to a unified description of the weak force and electromagnetism in a framework known as the ``electroweak'' theory. This process culminated in last quarter of the 20th century with the establishment of the Standard Model of particle physics which provides a unified - albeit, in some ways flawed - description of the weak, strong and electromagnetic forces as excitations of the quantum mechanical ``vacuum''. Gravity, however, remains outside the grasp of such unified frameworks. Consequently we are left in the awkward situation where the most complete formulation of physical law must be written in the form: Standard Model + Gravity.

\section{Superpositions of Geometry} 
\label{sec:superpositions}


The challenge in unifying gravity with quantum mechanics can be understood in the following way. The central feature of quantum mechanics is the principle of superposition - that the wavefunctions which describe two different particles or systems - can overlap. Consequently two systems described by two different wavefunctions $\Psi_1(x,t)$ and $\Psi_2(x,t)$ can instead be treated as a single composite system with wavefunction: $\Psi(x,t) = \Psi_1(x,t) + \Psi_2(x,t)$. Wavefunctions are just that - functions - and must therefore be defined on some set. In quantum mechanics the set is taken to be the co-ordinates $(x,t)$ of the spacetime our system is embedded in. So we can always write down what the wavefunction of a particle passing through two slits at the same time must look like. Similarly we can write down the superposition of two particles in terms of their momenta, rather than their position, by working in the $(p,t)$ basis instead of the $(x,t)$ basis with the momenta $p$ being related to the position $x$ by the usual fourier transform:
$$ \Psi(p,t) \sim \int dx \, e^{ipx} \Psi(x,t) $$
Regardless of whether we work in the position basis or the momentum basis, there is an implicit assumption at work in this prescription - this is the assumption of a flat background geometry for which we can assign a set of co-ordinates $(x,y,z,t)$ to each point of the spacetime.

Now, General Relativity teaches us that physics should be independent of the particular co-ordinates used to describe a system. Moreover, any theory which is consistent with GR must also be well-defined both on curved space as on flat space. It turns out that while we can perform quantum mechanical calculations to our heart's content with wavefunctions defined on flat space, the case of wavefunctions living on curved space becomes tricky. The complications associate with spacetime curvature can be dealt with by resorting to sufficiently sophisticated mathematical methods. The resulting framework is known as Quantum Field Theory on Curved Spacetime (QFT-CS), not a terribly memorable phrase. It was using the methods of QFT-CS that Stephen Hawking obtained his historic result \cite{Hawking1971Gravitational,Hawking1975Particle} that a black hole must emit thermal radiation at a rate inversely proportional to its mass. However, even QFT-CS does not qualify as a theory of ``quantum gravity'', as explained next.

Quantum Mechanics is about assigning various attributes to a system and then constructing \emph{states} of the system corresponding to each attribute. These states can then be superimposed and the resulting system will then manifest all non-intuitive phenomena associated with quantum behavior such as interference and entanglement. As mentioned previously, gravity in the modern conception arises from the non-trivial geometry of a region of spacetime induced by some distribution of matter. Some of the geometric attributes that one can assign to a given region of spacetime are \emph{length}, \emph{area}, \emph{volume}, \emph{angles} etc. A theory of quantum gravity should be able to tell us how to write down the wavefunction, not \emph{defined on} a given region of spacetime, but a wavefunction \emph{of} a given region of spacetime, allowing us to construct states which correspond to superpositions of different geometries. However as mentioned previously traditional quantum mechanics tells us only how to write down the wavefunction \emph{on} a given geometry rather than \emph{of} a given geometry. The traditional language of quantum mechanics is thus insufficient to describe quantum states of geometry.

For the same reasons QFT-CS is also not a theory of ``quantum gravity''. There the curved spacetime merely serves as an arena on which quantum states can be defined, but there is no notion of states of the geometry itself, rather than of the matter which moves about on that geometry.

\section{New Paradigms} 
\label{sec:new_paradigms}

At present there are several approaches towards tackling the open question of writing quantum states of geometry. These include String Theory \cite{Tong2010Lectures,Polchinski1998aString,Polchinski1998bString}, Loop Quantum Gravity \cite{Ashtekar2004Background,Dona2010Introductory,Rovelli2011Zakopane}, Causal Dynamical Triangulations \cite{Ambjorn2013Quantum,Loll2005The-Universe,Loll2008The-Emergence} among others. Most laypersons with an interest in science have heard of String Theory, simply because it is the oldest of these approaches and thus also the most widely taught and practiced. LQG was born about a decade after String Theory and has only recently reached a level of maturity and acceptability as a valid physical theory. It would take us far afield to go into details - even at a non-technical level - of these approaches and their similarities and differences. I will try to briefly summarize the two approaches and the basic idea behind each one.

The idea behind String Theory is that instead of a description of fundamental particles as point-like objects we should switch to a picture where the basic entities are extended one-dimensional objects called strings. These strings move and interact in some background spacetime. Requirements of physical and theoretical consistency restrict the number of dimensions of the spacetime in which strings can live to 26, 11, and 10 depending on the particular characteristics - fermionic, bosonic, open, closed - we choose to endow the strings with. The excitations of a string happen to include a part which can be identified with gravitons - which are excitations of the background geometry the string is propagating in. Though gravitons are often thought of as the quanta of the gravitational field, in the same way that photons are quanta of the electromagnetic field, this belief is only partially correct.

As mentioned in previous sections, the gravitational field is characterized by geometric attributes such lengths, areas and volumes. Therefore, quanta of the gravitational field should correspond to quantized lengths, areas and volumes, in the same way that a quantum of the electromagnetic field corresponds to a quantized amount of energy given by Planck's relationship between the energy of a photon and its frequency $E = h f$. However, the graviton picture does not predict any such relations between any fundamental geometric quantities - such as the area of a given region of spacetime and the frequency of a gravitational wave which passes through that region - and so cannot be said to provide a picture of quantum geometry. Moreover, gravitons are perturbations of the background spacetime which is, by default, presumed to be smooth and continuous. As such studying gravitons is analogous to studying the behavior of perturbations of a body of fluid. Studying the pertubations of a fluid will give us the theory of waves but will not inform us of the nature of the molecules and atoms which constitute the fluid. Similarly a study of gravitons allows us to study perturbations of the gravitational field but does not give us any indication of the ``molecules'' and ``atoms'' from whose combinations geometry - and therefore the gravitational field - arises.

LQG advocates a different perspective. From the very beginning\footnote{This statement is not quite accurate. A new avenue to approaching the problem of quantizing the gravitational field opened up after the introduction of the ``new variables'' by Abhay Ashtekar in 1987-88. It was only over the course of the next several years, work by Carlo Rovelli and Lee Smolin (among others) \cite{Rovelli1993Area,Rovelli1994Discreteness,Rovelli1995Spin} showed that following through with the quantization procedure in the new framework necessarily implied the existence of a discrete, quantized geometry at the Planck scale.} the notion of a smooth, continuous background geometry is abandoned in favor of a discrete geometry which is built out of elementary objects known as ``simplices'' - which is a complicated term for elementary geometric objects such as triangles and tetrahedra. In much the same way that Lego blocks can be glued together to build complicated structures, a collection of triangles or tetrahedra can be assembled to build a two-dimensional or three-dimensional geometry respectively. LQG allows us to calculate the quantized values of geometric attributes associated with these simplices. It provides us with a framework for studying \emph{quanta of geometry} - in the true sense of the phrase - and to construct superpositions of different states of geometry. However, there remain many shortcomings in LQG approach. Two significant obstacles are \emph{a).} the lack of a grasp on how we can obtain an (approximately) smooth, continuous spacetime by gluing together our elementary simplices and \emph{b).} a lack of understanding of how matter - particles such as electrons and neutrinos - should be described in terms of quanta of geometry.


\section{In Search of A New Unification} 
\label{sec:a_new_unification}

String Theory and LQG each have its own strengths and weaknesses. String Theory provides us with a description of matter in terms of extended stringy objects but does not address the question of the smoothness, or lack thereof, of spacetime. LQG provides us with a description of spacetime as being built out of ``atoms'' or quanta of geometry but does not tackle the question of describing matter degrees of freedom. Whatever form the final theory of quantum gravity takes, it is my personal belief\footnote{and \emph{my} personal belief is certainly \textbf{not} to be taken to be representative of what others in the research community might feel.}, that it will incorporate elements of both String Theory and LQG. Such a framework is not yet upon us, though we can see glimmering of its final shape. Moreoever, a complete understanding will certainly not be obtained by resorting only to those insights gained from research in high energy physics and ignoring the insights in other fields of physics such as the study of many body phenomena (known as ``condensed matter physics'') \cite{Vaid2013Non-abelian,Vaid2013Superconducting} or the field of quantum computation \cite{Vaid2013Elementary}.

Whatever the form the final description does take we are guaranteed a bonanza of new theoretical and experimental revelations in pursuit of the final theory. Apart from the sheer thrill of taking part in and completing the most recent stage of humanity's continuing quest to understand the inner workings of the Universe, there are also huge practical advantages to be gained from a complete and self-consistent theory of quantum gravity. When Newton developed his Laws of Motion and Gravity, did anyone forsee the technological developments which those laws would undergird over the course of the next three centuries? When Einstein developed relativity, both Special and General, did anyone forsee the myriad uses his breakthroughs would have in products such as GPS, graphene transistors and optical communications; not to mention the understanding we gained of less earthly phenomena such as the Big Bang and black holes? Similarly, we cannot even begin to fathom what riches an understanding of the properties of geometry and matter, under the umbrella of a theory of quantum gravity, will bring to our society and to the world at large.

Humanity has only just begun to venture out of the dark ages. Laptops, smart phones, fission, fusion and interplanetary (unmanned and soon enough, manned) missions only provide a glimpse of the wonders we are yet to uncover by harnessing those forces of Nature which as yet remain out of our grasp. The possibilities that lie ahead are truly ... limitless.


\bibliographystyle{JHEP}
\bibliography{qg_dummies.bib}

\end{document}